\documentstyle[ltwol,psfig]{article}

\def\bra{\langle}
\def\ket{\rangle}

\def\npb#1#2#3{    {\em Nucl. Phys. }{B\,#1} (19#2) #3}
\def\plb#1#2#3{    {\em Phys. Lett. }{B\,#1} (19#2) #3}
\def\prd#1#2#3{    {\em Phys. Rev. }{D\,#1} (19#2) #3}

\def\prl#1#2#3{    {\em Phys. Rev. Lett. }{#1} (19#2) #3}

\def\zpc#1#2#3{    {\em Zeit. f\"ur Physik }{C\,#1} (19#2) #3}

\def\beq{\begin{equation}}
\def\eeq{\end{equation}}
\def\bea{\begin{eqnarray}}
\def\eea{\end{eqnarray}}

\newcommand{\as}{\alpha_{\scriptscriptstyle S}}

\def\gtap{\ \raisebox{-.4ex}{\rlap{$\sim$}} \raisebox{.4ex}{$>$}\ }
\newcommand{\hd}{{\overline D}}

\newcommand{\m}{\mu}
\newcommand{\mb}{m_b}
\newcommand{\mw}{m_W}
\newcommand{\muw}{\mu_W}

\newcommand{\mub}{\mu_b}
\newcommand{\mh}{m_{H^\pm}}

\newcommand{\smallsm}{{\scriptscriptstyle SM}} 
\newcommand{\smallyy}{{\scriptscriptstyle YY}} 
\newcommand{\smallxy}{{\scriptscriptstyle XY}} 

\def\tanb{\mbox{$\tan \! \beta\,$}}

\newcommand{\bbxsg}{\mbox{$\overline{B} \to X_s \gamma\,$}}

\newcommand{\bbrx}{\mbox{${\rm BR}(\overline{B}\to X_s\gamma)\,$}}

\newcommand{\xx}{\mbox{$\rm{X}$}}
\newcommand{\yy}{\mbox{$\rm{Y}$}}

\begin{document}


\begin{titlepage}

\begin{flushright}
PM--98/23            \\
BUTP--98/26          \\[3ex] 
\end{flushright}
\vspace{2.5cm}

\begin{center}
{\bf LESSONS FROM $\boldmath{\overline{B} \to X_S \gamma}$ 
 IN TWO HIGGS DOUBLET MODELS}                       \\[5ex] 
 F. M.\ BORZUMATI                                   \\[1ex]
{\it Laboratoire de Physique Math\'ematique et Th\'eorique, 
     Universit\'e Montpellier II,}                  \\ 
{\it F-3405 Montpellier Cedex 5, France}            \\[3ex]
 C. GREUB                                           \\[1ex]
{\it Institut f\"ur theoretische Physik, 
     Universit\"at Bern},                           \\  
{\it Sidlerstrasse 5, 3012 Bern, Switzerland}       \\[10ex]
\end{center}
\begin{center} 
ABSTRACT \\
\vspace*{1mm}
\parbox{13cm}{
The next--to--leading order predictions for the branching
ratio ${\rm BR}(\overline{B} \to X_s \gamma)$
are given in a  
generalized class of two Higgs doublets models. Included are the
recently calculated leading QED corrections. It is shown that the high
accuracy of the Standard Model calculation is in general not shared by
these models. Updated lower limits on the mass of the charged Higgs
boson in Two Higgs Doublet Models of Type~II are presented.
}
\end{center}

\vspace*{2truecm}
{\begin{center} 
{\it Talk presented by C. Greub at the } \\
{\it XXIX International Conference on High Energy Physics}      \\
{\it Vancouver, B.C., Canada, 23--29 July 1998}
\end{center}}

\vfill
\end{titlepage}

\thispagestyle{empty}



\title{LESSONS FROM $\boldmath{\overline{B} \to X_S \gamma}$ 
 IN TWO HIGGS DOUBLET MODELS}

\author{F.M. BORZUMATI}

\address{Laboratoire de Physique Math\'ematique et Th\'eorique, 
 Universit\'e Montpellier II, F--3405 Montpellier Cedex 5, France\\
 E-mail: francesc@lpm.univ--montp2.fr}   

\author{C. GREUB}

\address{Institut f\"ur theoretische Physik, Universit\"at Bern, 
 Sidlerstrasse 5, 3012 Bern, Switzerland
\\E-mail: greub@itp.unibe.ch}  


\twocolumn[\maketitle\abstracts{
The next--to--leading order predictions for the branching
ratio ${\rm BR}(\overline{B} \to X_s \gamma)$ are given in a  
generalized class of two Higgs doublets models. Included are the
recently calculated leading QED corrections. It is shown that the high
accuracy of the Standard Model calculation is in general not shared by
these models. Updated lower limits on the mass of the charged Higgs
boson in Two Higgs Doublet Models of Type~II are presented.
}]

\section{Introduction}
\label{sec:introduction}
\noindent 
Two Higgs Doublet Models (2HDMs) are conceptually among the simplest
extensions of the Standard Model (SM). They contain additional sources
of flavour change due to their extended Higgs sectors. Studies of the
$\bbxsg$ decay in this class of models, therefore, can already test
how unique is the accuracy of the SM result for the branching ratio
\bbrx at the next--to--leading (NLO) order in QCD~\cite{BG}, even
requiring calculations at the same level of complexity as the SM
one. They can obviously provide also important indirect bounds on the
new parameters contained in these models. In spite of their apparent
simplicity, indeed, 2HDMs have not been correctly constrained in
ongoing experimental searches~\cite{KRA,BD}.

The well--known Type~I and Type~II models are particular examples of
2HDMs, in which the same or the two different Higgs fields couple to
up-- and down--type quarks.  The second one is especially important
since it has the same couplings of the charged Higgs $H^+$ to fermions
that are present in the Minimal Supersymmetric Standard Model
(MSSM). The couplings of the neutral Higgs to fermions, however, have
important differences from those of the MSSM~\cite{KRA,BD}. Since,
beside the $W$, only charged Higgs bosons mediate the decay $\bbxsg$
when additional Higgs doublets are present, the predictions of $\bbrx$
in a 2HDM of Type II give, at times, a good approximation of the value
of this branching ratio in some supersymmetric models~\cite{MMM}.

It is implicit in our previous statements that we do not consider
scenarios with tree--level flavor changing couplings to neutral Higgs
bosons.  We do, however, generalize our class of models to accommodate
Multi--Higgs Doublet models, provided only one charged Higgs boson
remains light enough to be relevant for the process $\bbxsg$.  This
generalization allows a simultaneous study of different models,
including Type~I and Type~II, by a continuous variation of the
(generally complex) charged Higgs couplings to fermions.  It allows
also a more complete investigation of the question whether the
measurement of $\bbrx$ closes the possibility of a relatively light
$H^\pm$ not embedded in a supersymmetric model.

This summer (1998), a new (preliminary) measurement of this decay rate
was reported by the CLEO Collaboration~\cite{CLEOneu}
$ \bbrx = (3.15 \pm 0.35\pm 0.32 \pm 0.26)\times 10^{-4}\,,$
which, compared to older results, is based on $53\%$ more data ($3.3
\times 10^6$ $B \bar{B}$ events).  The upper limit allowed by this 
measurement, reported in the
same paper, is $4.5\times 10^{-4}$ at $95\%$ C.L..  The ALEPH
Collaboration has measured~\cite{ALEPHneu}
$ \bbrx =(3.11\pm 0.80\pm0.72)\times 10^{-4} $
from a sample of b hadrons produced at the $Z$--resonance.

The theoretical prediction for \bbrx has a rather satisfactory level
of accuracy in the SM. The main uncertainty, slightly below $\pm10\%$,
comes from the experimental error on the input parameters.  The more
genuinely theoretical uncertainty, due to the unknown value of the
renormalization scale $\mub$ and the matching scale $\muw$, which was
unacceptably large at the
leading--order (LO) level, was reduced to roughly $\pm 4\%$ when the 
NLO QCD corrections to the partonic decay width 
$\Gamma(b \to X_s \gamma)$ were completed~\cite{common}.
(See ref.~\cite{BG} for the milestone papers which brought
to the complete LO calculation.) 
In addition,
non--perturbative contributions to $\bbrx$, scaling like
$1/m_b^2$~\cite{mbcorr} and $1/m_c^2$~\cite{mccorr}, were
computed. Very recently, the leading QED and some classes 
of electroweak corrections were also 
calculated~\cite{Marciano,Strumia,Kagan}. 

Following the procedure described in refs~.\cite{BG,newcontour},  
and including QED corrections as in ref.~\cite{Kagan}, 
we obtain a theoretical prediction~\cite{newcontour} in agreement 
with the existing data: 
\beq
\label{end2}
 \bbrx = (3.32 \pm 0.14 \pm 0.26)\times 10^{-4}.
\eeq
The first error in (\ref{end2}) is due to the $\mub$ and $\muw$ scale 
uncertainties; the second, comes from the experimental 
uncertainty in the input parameters. 

A detailed study of $\bbxsg$ at the NLO in QCD~\cite{BG} in 2HDMs, on
the contrary, shows that the NLO corrections and scale dependences in
the Higgs contributions to $\bbrx$ are very large,
irrespectively of the value of the charged Higgs couplings to
fermions. This feature remains undetected in Type~II models, where the
SM contribution to $\bbrx$ is always larger than, and in phase with,
the Higgs contributions. In this case, a
comparison between theoretical and experimental results for $\bbrx$
allows to conclude that values of $\mh = O(\mw)$ can be
excluded. Such values are, however, still allowed in other
2HD models.

These issues are illustrated in Sec.~4, after defining in Sec.~2 the
class of 2HDMs considered, and presenting the NLO corrections at the
amplitude level in Sec.~3. A brief discussion on the existing lower
bounds on $\mh$ coming from direct searches at LEP is included in 
Sec.~3. This rests on observations brought forward in 
refs.~\cite{KRA,Maettig} contributed to this conference. 

%
\section{Couplings of Higgs bosons to fermions}
\label{2hdms}

\noindent 
Models with $n$ Higgs doublets have generically a Yukawa 
Lagrangian (for the quarks) of the form:
\beq 
-{\cal L} =
  h^d_{ij} \,{\overline{q'}_L}_i \,\phi_1 \, {d'_R}_j 
 +h^u_{ij} \,{\overline{q'}_L}_i \,{\widetilde \phi}_2 \,{u'_R}_j 
 +{\rm h.c.}\,,
\label{yukpot}
\eeq
where $q'_L$, $\phi_i$, ($i=1,2$) are SU(2) 
doublets (${\widetilde \phi_i} = i \sigma^2 \phi_i^*$); 
$u'_R$, $d'_R$ are SU(2) singlets
and 
$h^d$, $h^u$ denote $3\times3$ Yukawa matrices. 
To avoid 
flavour changing neutral couplings at the 
tree--level, it is sufficient to 
impose that no more than one Higgs doublet couples to the same 
right--handed field, as in eq. (\ref{yukpot}). 

After a rotation of the quark fields from the current eigenstate to
the mass eigenstate basis, and an analogous rotation of the charged
Higgs fields through a unitary $n \times n$ matrix $U$, we assume 
that only one of the $n-1$ charged physical Higgs bosons
is light enough
to lead to observable effects in low energy processes. 
The $n$--Higgs doublet model then reduces to a generalized
2HDM, with 
the following Yukawa interaction for this 
charged physical Higgs boson denoted by $H^+$:
\beq 
{\cal L} =\frac{\,g}{\sqrt{2}} \left\{
\frac{{m_d}_i}{\mw}
      \xx \,{\overline{u}_L}_j V_{ji}  \, {d_R}_i+
\frac{{m_u}_i}{\mw}
      \yy \,{\overline{u}_R}_i V_{ij}  \, {d_L}_j
                               \right\} H^+ \,.
\label{higgslag}
\eeq
In~(\ref{higgslag}), $V$ is the Cabibbo--Kobayashi--Maskawa matrix
and the symbols $\xx$ and $\yy$
are defined in terms of elements of 
the matrix $U$ (see citations in ref. \cite{BG}). 
Notice that $\xx$ and $\yy$ are
in general complex numbers and therefore potential sources of
CP violating effects. 
The ordinary Type~I and Type~II 2HDMs (with $n=2$), are
special cases of this generalized class, with 
$(\xx,\yy) = (- \cot \beta, \cot \beta)$
and
$(\xx,\yy) = ( \tan \beta, \cot \beta)$, respectively.

We do not attempt to list here the generic couplings of fermions to
neutral Higgs fields. In a 2HDM of Type~II, the 
neutral physical states are two CP--even states $ h$ and $H$ 
($m_h < m_H$) and a CP--odd state $A$. In this case, only one additional
rotation matrix is needed, parametrized by the 
angle $\alpha$, which is independent of the rotation
angle $\beta$ of the charged sector. This independence 
stops to be true when this model 
is supersymmetrized since supersymmetry induces a
relation between $\tan 2 \alpha$ and $\tan 2 \beta$.


\section{NLO corrections at the amplitude level}
\noindent 
The NLO corrections are calculated using 
the framework of an effective low--energy theory with five
quarks, obtained by integrating out the
the $t$--quark, the $W$--boson and 
the charged Higgs boson. The relevant effective 
Hamiltonian for radiative $B$--decays 
\beq
\label{heff}
 {\cal H}_{eff} = - \frac{4 G_F}{\sqrt{2}} \,V_{ts}^\star V_{tb} 
   \sum_{i=1}^8 C_i(\mu) {\cal O}_i(\mu)  
\eeq
consists precisely of the same operators ${\cal O}_i(\mu)$
used in the SM case, weighted by the Wilson coefficients 
$C_i(\mu)$. The explicit form of the operators can be seen 
elsewhere~\cite{BG}.

Working to NLO precision means that one is resumming all the
terms of the form $\as^n(\mb) \, \ln^n (\mb/M)$, as well as
$\as(\mb) \, \left(\as^n(\mb) \, \ln^n (\mb/M)\right)$. The symbol
$M$ stands for one of the heavy masses $\mw$, $m_t$ or $m_H$ which 
sets the order of magnitude of the matching scale $\muw$.
This resummation is achieved through the following 3 steps: 
{\it 1)} 
matching the full standard model theory with the effective theory at
the scale $\muw$.  The Wilson coefficients are needed at the $O(\as)$
level;
{\it 2)} 
evolving the Wilson coefficients from $\m=\muw$ down to $\m = \mub$,
where $\mub$ is of the order of $m_b$, by solving the appropriate
renormalization group equations.  The anomalous dimension matrix has
to be calculated up to order $\as^2$;
{\it 3)} 
including corrections to the matrix elements of the operators 
$\langle s \gamma | {\cal O}_i (\mu) |b \rangle$ at the scale 
$\mu = \mub$ up to order $\as$.

Only step {\it 1)} gets modified when including the
charged Higgs boson contribution to the SM one.
The new contributions to the matching conditions have been
worked out independently by several groups~\cite{CRS,CDGG,BG}, 
by simultaneously integrating out 
all heavy particles, $W$, $t$, and $H^+$ at the scale
$\muw$. This is a reasonable 
approximation provided $\mh$ is of the same order of magnitude
as $\mw$ or $m_t$.     

Indeed, the lower limit on $\mh$ coming from LEP~I, of 
$45\,$GeV, guarantees already $\mh = O(\mw)$. There 
exists a higher lower bound from LEP~II of 
$55\,$GeV for any value of $\tan \beta$~\cite{JANOT} 
for Type~I and Type~II models, which has been recently criticized in 
ref.~\cite{BD}. This criticism is based on the fact that 
there is no lower
bound on $m_A$ and/or $m_h$ coming from LEP~\cite{KRA}.
As it was already mentioned, in 2HDMs (unlike in the MSSM),
the two rotation angles of the neutral and charged Higgs 
sector, $\alpha$ and $\beta$, are independent parameters. 
Therefore, the pair--production
process $e^+e^- \to Z^\ast \to h A$ and the 
Bjorken process $e^+e^- \to Z^\ast \to Z h$, which are 
sufficient in the MSSM to put lower limits on 
$m_h$ and $m_A$ separately, imply in this case only that
$m_h+m_A \gtap 100\,$GeV~\cite{Desch}.
The other two production mechanisms possible at LEP~I (they require
larger numbers of events than LEP~II can provide) are the decay 
$Z \to h/A\gamma$ and the radiation of $A$ off $b \bar b$ and 
$\tau^+\tau^-$ pairs. The latter, allows 
for sizable rates only for large values of $\tan\!\beta$ and it 
yields the constraint $\tan \beta \le 40$ for $m_A =15$ GeV, in a 
2HDM of Type~II~\cite{PA13-027}. 
The former one, limits weekly $\tan\beta$ to be in the range 
$\{0.2, 100\}$ for $m_{h/A} \approx 10$ GeV~\cite{Maettig}.
Indirect searches from the anomalous magnetic moment of the muon also
lead to constraints on $\tan \beta$ for a light $h$ and $A$.
For a pseudoscalar this limit is stronger than that 
obtained from the radiation mechanism at LEP~I 
only for $m_{A} < 2$ GeV~\cite{KRA}. 

Therefore, one of the neutral Higgs bosons can still be light. Charged
Higgs bosons pair--produced at LEP~II can then decay as
$ H^+ \to A W^+$ and/or $ h W^+$ with an off--shell $W$ boson. The 
rate is not negligible and invalidates the limit of $55\,$GeV
in Type~I models and models of Type~II with 
modest $\tan \beta$, obtained considering only 
$c \bar{s}$ and $\tau^+ \nu_\tau$ as possible decay modes 
of $H^\pm$. The unescapable limit of LEP~I is 
however already large enough for a simultaneous integration out
of $H^+$, $W$ and $t$. 

After performing steps {\it 1)}, {\it 2)}, and {\it 3)}, it is 
easy to obtain the quark level amplitude
$A(b \to s \gamma)$.
As  the matrix elements $\bra s \gamma|{\cal O}_i|b \ket$ are proportional
to the tree--level matrix element of the operator ${\cal O}_7$,
the amplitude $A$ can be written in the compact form
\beq
\label{ampl}
A(b \to s \gamma) = \frac{4 G_F}{\sqrt{2}} V_{tb} V_{ts}^* \, \hd \,
\bra s \gamma|{\cal O}_7|b\ket_{tree} \quad .
\eeq
(It should be noticed that a subset of Bremsstrahlung
contributions was transferred to (\ref{ampl}), as described in 
ref.~\cite{BG}.)
For the following discussion it is useful to decompose the reduced amplitude 
$\hd$
in such a way that the dependence on the couplings $\xx$ and $\yy$
(see eq. (\ref{higgslag}))
becomes manifest:
\beq
\label{dsplit}
\hd = \hd_\smallsm + \xx \yy^* \hd_\smallxy + |\yy|^2 \hd_\smallyy \quad .
\eeq
\begin{figure}
\center
\psfig{figure=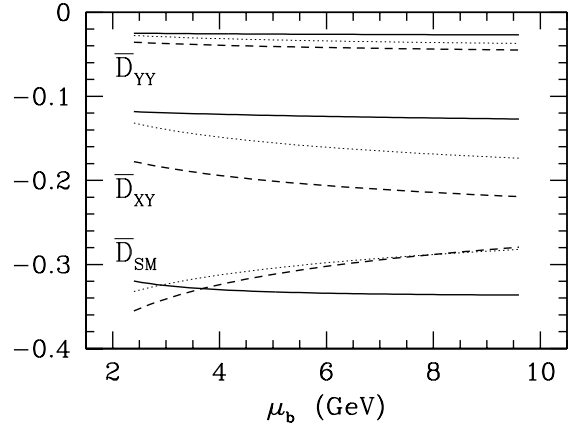,height=2.16in}
\caption{LO (dashed) and NLO (solid) predictions of the various pieces
of the reduced amplitude $\hd$ for $\mh =100\,$GeV (see text).}
\label{fig:radk}
\end{figure}
In Fig. 1 
the individual $\hd$ quantities are shown in LO (dashed) and NLO (solid) 
order, for $\mh = 100$ GeV, as a function of $\mu_b$; 
all the other input parameters
are taken at their central values, as specified in ref. \cite{BG}.  
To explain the situation, one can concentrate on the curves for 
$\hd_\smallxy$.
Starting from the LO curve (dashed), the final NLO prediction is due
to the change of the Wilson coefficient $C_7$, shown by 
the dotted curve, and by the inclusion of the virtual
QCD corrections to the matrix elements. This results into a further
shift from the dotted curve to the solid curve. Both effects
contribute with the same sign and with similar magnitude, as it 
can be seen in Fig.~1. 
The size of the NLO corrections in the term $\hd_\smallxy$ in
(\ref{dsplit}) is
\beq
\frac{\Delta \hd_\smallxy}{\hd_\smallxy^{LO}} \equiv
\frac{\hd_\smallxy^{NLO}-\hd_\smallxy^{LO}}{\hd_\smallxy^{LO}}
\sim - 40\% \,!
\eeq
A similarly large correction is also obtained for
$\hd_\smallyy$. For the SM contribution $\hd_\smallsm$,
the situation is different: the corrections to the
Wilson coefficient $C_7$ and the corrections due to the virtual
corrections in the matrix elements are smaller individually, 
and furthermore tend to cancel when combined, as shown in Fig. 1.

The size of the corrections in $\hd$ strongly depends on the 
couplings $\xx$ and $\yy$ (see eq. (\ref{dsplit})):
$\Delta \hd/\hd$ is small, if the SM dominates, but it 
can reach values such as $-50\%$ or even worse, if the SM and 
the charged Higgs
contributions are similar in size but opposite in sign. 

\section{Results and Conclusions}
\begin{figure}
\center
\psfig{figure=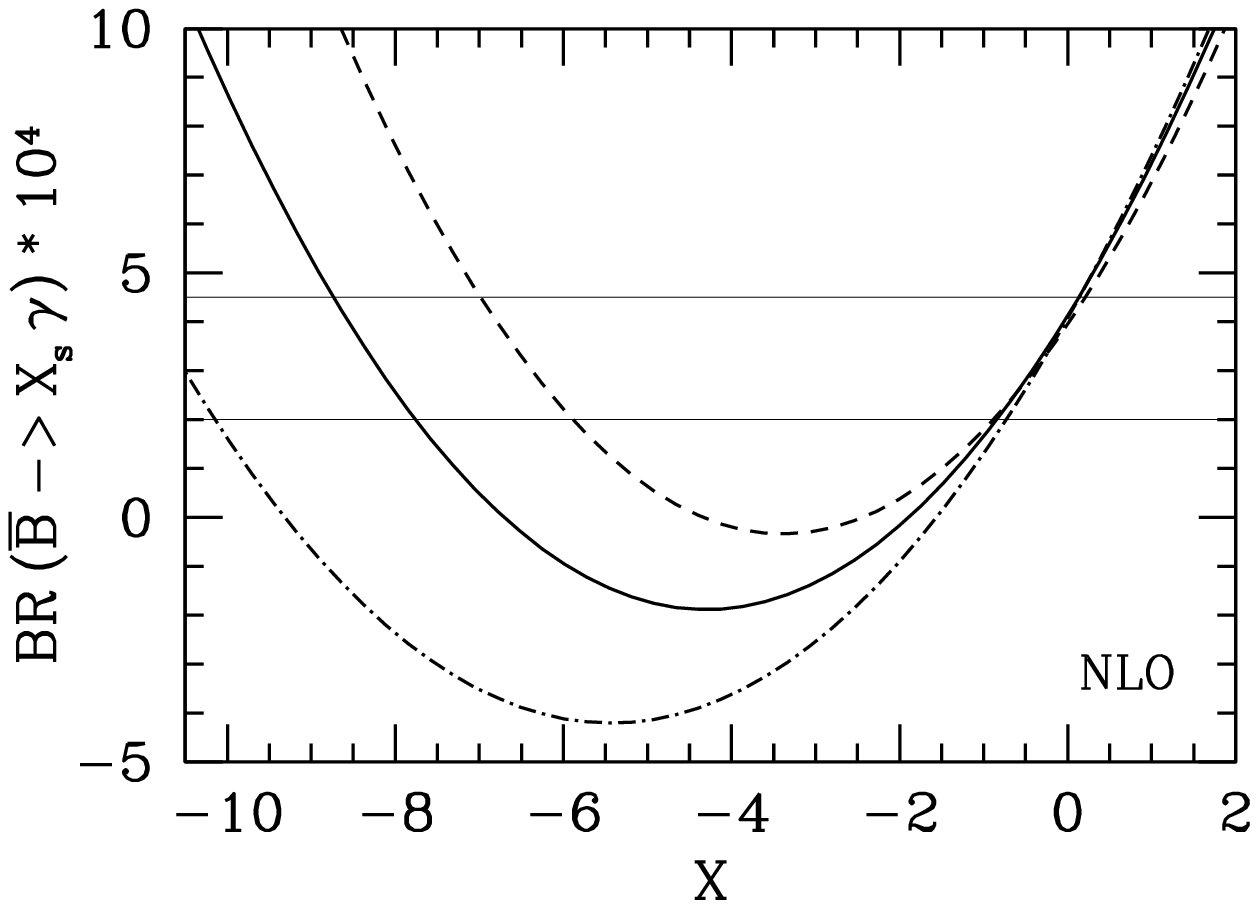,height=2.16in}
\caption{\bbrx for $\yy=1$, $\mh=100$ GeV as a function of $\xx$,
for $\mub=4.8$ GeV (solid), $\mub=2.4$ GeV (dahed) and
$\mub=9.6$ GeV (dash-dotted). 
Superimposed is the range of values allowed by the CLEO measurement.}
\end{figure}
\noindent
The branching ratio \bbrx can be schematically written as
\beq
\label{schematic}
\bbrx  \propto |\hd|^2 + \cdots 
\quad ,
\eeq
where the ellipses stand for Bremsstrahlung contributions, electroweak
corrections and non--perturbative effects.
As required by perturbation theory,
$|\hd|^2$ in eq. (\ref{schematic}) should
be understood as
\beq
\label{dsq}
|\hd|^2 = |\hd^{LO}|^2 \left[ 1 + 2 \mbox{Re} \left( 
\frac{\Delta \hd}{\hd^{LO}} \right)
\right] \quad ,
\eeq
i.e., the term $|\Delta \hd/\hd^{LO}|^2$ is omitted.
If $\mbox{Re}(\Delta \hd/\hd^{LO})$ is larger than $50\%$ in magnitude
and negative, the NLO branching ratio becomes  negative, i.e. the 
truncation of the perturbative series at the NLO level is not 
adequate for the corresponding couplings $\xx$ and $\yy$.
This happens also for modest 
values of $\xx$ and $\yy$, as it is illustrated in Fig.~2, where 
only real couplings are considered.
The values $\xx=1$ and $\xx=-1$ in this figure, correspond 
respectively to the predictions of a the Type~II
and a Type~I 2HDM with $\tan \beta =1$.  

Theoretical predictions for the branching ratio in 
Type~II models stand, in general, on a rather solid ground.
Fig.~3 shows the low--scale 
dependence of $\bbrx$ for matching scale $\muw = \mh$, for 
$\mh>100\,$GeV. It is less than $\pm 10\%$ for any value of 
$\mh$ above the LEP~I lower bound of $45\,$GeV. Such a small scale 
uncertainty 
is a generic feature of Type~II models and remains true for 
values of $\tanb$ as small as $0.5$.
In this, as in the following figures where reliable NLO 
predictions 
are presented, the recently calculated leading QED 
corrections are included in the way
discussed in the addendum~\cite{newcontour} of
ref.~\cite{BG}. They are not contained in the result
shown in Figs.~1 and 2, which have only an illustrative 
aim. 

In Type~II models, the theoretical estimate of $\bbrx$ can be well 
above the experimental upper bound of $4.5 \times 10^{-4}$ ( $95\%$ C.L.), 
reported by the CLEO Collaboration at this conference~\cite{CLEOneu}, 
leading to constraints in the $(\tan \beta, \mh)$
plane. The region excluded by the 
CLEO bound, as well as by other   
hypothetical experimental bounds, is given in Fig. 4.
These contours are obtained 
minimizing the ratio 
${\rm BR}(\overline{B}\to X_s\gamma)/
{\rm BR}(b \to c l \nu_l)$, when varying simultaneously the 
input parameters within their errors as well as the two scales
$\mub$ and $\muw$.
For $\tanb = 0.5,1,5$, we exclude respectively 
$\mh \le 280$,$\,200$,$\,170\,$ GeV, using the present upper bound from
CLEO.
Notice that the flatness of the curves shown in Fig.~3
towards the higher end of $\mh$, 
causes a high sensitivity of these bounds on all details 
of the calculation (see ref.~\cite{BG}). 
These details can only alter the 
branching ratio at the $1\%$ level, i.e. well within the 
estimated theoretical uncertainty, but they may 
produce shifts of several 
tens of GeV, in either direction, in the lower bounds quoted 
above.    
\begin{figure}
\center
\psfig{figure=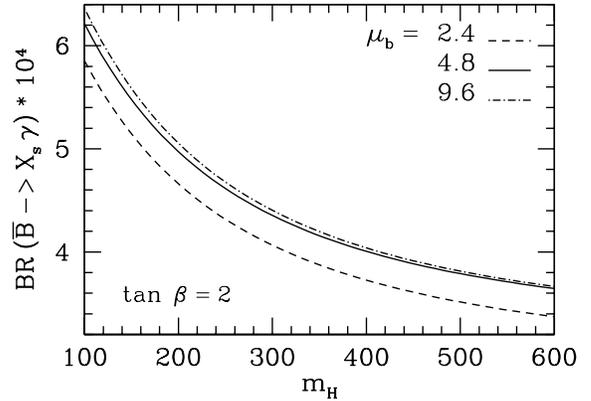,height=2.16in}
\caption{\bbrx in a Type~II model with $\tan \beta =2$, for
various values of $\mub$. The leading QED corrections are
included (see text).}
\end{figure}
\begin{figure}
\center
\psfig{figure=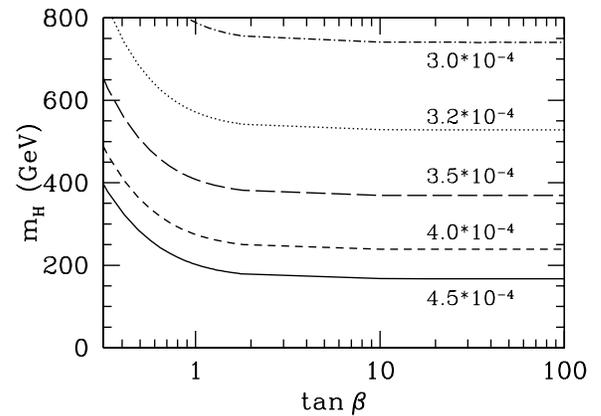,height=2.16in}
\caption{Contour plot in $(\tan \beta,\mh)$ in Type~II models, 
obtained by using the
NLO expression for \bbrx and possible experimental upper bounds.
The leading QED corrections are also included (see text). The 
allowed region is above the corresponding curves.}
\end{figure}

Also in the case of complex couplings, the results for 
\bbrx range from ill--defined, to uncertain, up to 
reliable. One particularly interesting case in which 
the perturbative expansion can be safely truncated at the NLO level,
is identified by: 
$\yy=1/2$, $\xx = 2 \exp (i\phi)$,  and $\mh=100\,$GeV. 
The corresponding branching ratios, shown in Fig.~5, 
are consistent with the 
CLEO measurement, even for a relatively small value of $\mh$
in a large range of $\phi$. 
Such a light charged Higgs can contribute to 
the decays of the $t$--quark, through the mode $t \to H^+ b$.
\begin{figure}
\center
\psfig{figure=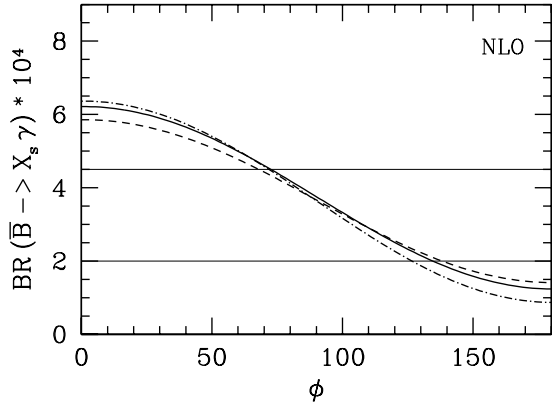,height=2.16in}
\caption{\bbrx as a function of $\phi$, where $\phi$ parametrizes
$\xx=2 \exp(i\phi)$, for $\yy=1/2$, $\mh=100$ GeV. Solid, dashed
and dash--dotted lines correspond to $\mub=4.8,2.4,9.6$ GeV.
The leading QED corrections are included (see text).
Superimposed is the range of values allowed by the CLEO measurement.}
\end{figure}

The imaginary parts in the $\xx$ and 
$\yy$ couplings induce --together with the absorptive parts
of the NLO loop-functions-- CP rate asymmetries in \bbxsg.
A priori, these can be expected to be large. We find, however, 
that choices of the couplings $\xx$ and $\yy$ which render the 
branching ratio stable, induce in general small asymmetries, not
much larger than the modest value of $1\%$ obtained in the SM. 

We conclude with the most important lessons which can be extracted out
of the calculation presented here.  The high accuracy reached in the
theoretical prediction of \bbrx at the NLO for the SM, is not a
general feature.  In spite of its potential sensitivity to new sources
of chiral flavour violation, the \bbxsg decay may turn out to be
unconstraining for many models, because of the instability of NLO
calculations. This is not only a temporary situation, since it is
higly unlikely that a higher order QCD improvement is carried out.
Nevertheless, there are scenarios in which \bbrx can be reliably
predicted at the NLO level as in 2HDMs of Type~II. In these models,
$\mh$ can then be safely excluded up to values which depend on the
experimentally maximal allowed value of \bbrx. Today, we find 
$\mh \gtap 165\,$GeV.  It has to be stressed, however, that in the
more general 2HDMs discussed in this article, $H^+$ can be much
lighter and may still be detected as a decay product of the
$t$--quark.

\section*{Acknowledgements}
We thank M. Krawczyk for helpful discussions.
This work was partially supported by CNRS and the Swiss National
Foundation.

\end{document}